\documentclass[aps,twocolumn,floatfix]{revtex4}
\usepackage{amssymb,amsmath,graphicx}
\begin{document}
\title{The Lindley paradox in optical interferometry}
\author{Camillo Mauri}
\affiliation{Quantum Technology Lab, Dipartimento di Fisica, Universit\`a 
degli Studi di Milano, I-20133 Milano, Italia}
\author{Matteo G. A.  Paris}\email{matteo.paris@fisica.unimi.it}
\affiliation{Quantum Technology Lab, Dipartimento di Fisica, Universit\`a 
degli Studi di Milano, I-20133 Milano, Italia}
\affiliation{CNISM, Unit\'a` Milano Statale, I-20133 Milano, Italia}
\affiliation{INFN, Sezione di Milano, I-20133 Milano, Italia}
\begin{abstract}
The so-called {\em Lindley paradox} is a counterintuitive statistical
effect where the Bayesian and frequentist approaches to hypothesis 
testing give radically different answers, depending on the choice of 
the prior distribution. In this paper we address the occurrence of the 
Lindley paradox in optical interferometry and discuss its implications 
for high-precision measurements. In particular, we focus on 
phase estimation by Mach-Zehnder interferometers and show how to 
mitigate the conflict between the two 
approaches by using suitable priors. 
\end{abstract}
\date{\today}
\maketitle
\section{Introduction}
\label{intro}
Interferometric setups are at the heart of several high-precision
measurement schemes, ranging from gravitational wave-detectors 
to laser gyroscopes and clocks synchronization protocols. 
In particular, a direct detection of gravitational waves is supposed 
to come from laser-interferometric detectors \cite{rev80,rev14,rev08}, 
where they induce a measurable variation of the optical paths of the light
beams traveling along the arms of an interferometric setup. 
\par
Inteferometers are usually employed to detect small perturbations, 
so that detectors must be able to measure distance changes with
remarkable precision and any source of noise must be carefully 
removed. Indeed, recent technological advances in precision lasers,
vacuum technology and optical systems had made it possible to greatly
reduce classical noise.  However, an unavoidable constraint to the
precision with which optical signals can be measured arises from the
quantum nature of the electromagnetic field. The so-called standard
quantum limit (SQL) \cite{bra92}, sometimes also referred to as {\em
shot noise}, results as a consequence of the uncertainty relations
existing for the quantum field operators. 
\par
A way to partially circumvent the effects of quantum fluctuations and
increase interferometric precision is to exploit the use of squeezed
light \cite{cav81,bon84,yur86} or other nonclassical states of light
\cite{hol93,sun08}, also in the presence of inefficient detectors
\cite{par95}. The implementation of squeezing-enhancement techniques in
interferometric detectors has been a great challenge: after a large number of
experimental and theoretical studies it has now became possible to take
advantage of squeezing \cite{geo600,berni15} so that the next 
generation of interferometers will be equipped with quantum technologies.  
\par
In order to effectively exploit the potential improvements offered by
quantum signals, the statistical analysis of the data should be refined
as well. In this context, the key point is to improve the ability of
discriminating whether a signal contains instrumental noise only or also a
trace of a signal. This problem, which arises in the form of a
hypothesis testing problem \cite{par97}, might be conveniently studied
by means of Bayesian inference
\cite{hol93,pez07,pez08,oli09,ber09,gen11,gen12,oli13}.  However, a Bayesian
approach may be challenging since it requires the full knowledge of the
system under consideration in order to properly introduce a suitable
prior, together with a high computational power. Indeed, in a Bayesian
framework we have to provide explicit models of the signal induced by 
a perturbation to achieve, after specifying our prior knowledge, 
the odds ratio of
the detection hypothesis over the null hypothesis, corresponding to
absence of signal. In this context, it may happen that the lack of
knowledge about the system increases the false alarm probability. In
other words, the posterior odds may favor the alternative hypothesis, i.
e. presence of perturbations, even if no signals are actually 
present in data. In
particular, when a diffuse prior distribution is taken, a Bayesian
approach to hypothesis testing leads to conflicting evidences compared
to a frequentist approach.  Such a counterintuitive situation was 
studied by Jeffrey \cite{jef48} together with Lindley, who first
referred to this disagreement as a statistical paradox \cite{lin57}. 
\par
The relevance of Bayesian analysis in interferometry has been recognized
since a long time \cite{hol93,hra96}. The interest raised for at least
three reasons. On the one hand, common interferometric estimators are
known to achieve optimal performances only at a specific working point
\cite{cav81,dow98}, thus resulting in interferometric protocols that do
not allow the measurement of arbitrary phase shifts.  Besides, in order
to estimate small fluctuations around the working point, the
interferometer has to be actively stabilized  by the addition of some
feedback mechanism \cite{fed96}.  Finally, Bayesian estimators have been
shown to achieve the asymptotic regime, where they saturate the
Cramer-Rao bound, already with few measurements
\cite{berni15,oli09,ber09,gen11}, thus representing a convenient choice
in any setting where resources are limited or the samples involved in
the interferometric setup are fragile.
\par
In this work we address in details the occurrence of the {\em Lindley
paradox} \cite{lin77,sha82} in the analysis of data coming from optical
interferometry. The importance of the topic seats in the fact that
often, e.g. in the framework of gravitational antennas, each event is of
great relevance and should be analyzed in the most refined way as the
above mentioned debate concerning cosmological measurements suggests.
More generally, it is often the case that besides the estimation of
small fluctuations, interferometric measurements are involved in
monitoring a given physical configuration, where the information coming
from the experimental data is exploited to statistically discriminate
between two hypothesis \cite{wis09,kum11,rei11,kir14,fen14}.  Here, we
first consider homodyne detection, then classical-like Mach-Zehnder (MZ)
interferometer, and finally we consider the case of squeezed-enhanced
quantum interferometers. For the MZ interferometer we start from the
analysis of the ideal case wherein the detectors are perfectly
efficient, and a widespread prior is assumed, and then proceed by
assessing the realistic case where inefficient detectors and a
concentrated prior distribution is employed.  As we will see, the
Lindley paradox may indeed arise in optical interferometry. Here we show
how to mitigate the conflict between the Bayesian and frequentist
approaches by using suitable priors. 
\par
The paper is organized as follows. In Sec. \ref{s:Lin} we introduce the
notation and illustrate how and when the paradox can arise. In Sec.
\ref{s:HD} we discuss phase estimation by homodyne
detection as a suitable example to illustrate the statistical
analysis step by step.  In Sec. \ref{s:MZI} we analyze in details the
occurrence of the paradox in optical interferometry, whereas Section
\ref{s:outro} closes the paper with some concluding remarks.
\section{The Lindley Paradox}
\label{s:Lin}
Let us consider a random variable $X$ distributed according to a
Gaussian distribution of unknown mean $\phi$ and known 
variance $\sigma^{2}$. Suppose
that we want to test a sharp null hypothesis $H_{0}$, corresponding to
the prediction $\phi=\phi_{0}$, against the alternative hypothesis
$H_{1}$, corresponding to the diffuse prediction $\phi\neq\phi_{0}$.  In
a Bayesian framework this is done assigning a priori a probability
$p\left(H_{i}\right)$ and a prior distribution
$\pi_{i}\left(\phi\right)$ to any
hypothesis $H_{i}$ taken into account. Let us denote the priors by
$p\left(H_{0}\right)=z_{0}$ and $p\left(H_{1}\right)=1-z_{0}$, with
$z_0\in [0,1]$. Besides, we take $\pi_{0}\left(\phi\right)$ to be a
Dirac's delta function since the null hypothesis involves a single
parameter value. As for the probability density $\pi_1 (\phi)$, 
concerning the alternative hypothesis, we assume a normal distribution 
with variance $\tau^{2}$. That being said, the posterior probability
$\bar{z} _{0}$ that the outcome $X=x$ provides evidence of
$\phi=\phi_{0}$ , i. e. confirms the null hypothesis, is evaluated via
Bayes' theorem, which yields (see \ref{a:pd}):
\begin{equation}
\bar{z} _{0} = p \left( \phi=\phi_{0} | x \right) =  
\left\{ 1 + \frac{ 1 - z_{0}}{z_{0}} \frac{ p \left( x| \phi 
\neq \phi_{0}\right) }{p \left( x | \phi=\phi_{0} \right) } \right\}
^{-1}\,.
\label{eq:1}
\end{equation}
In order to illustrate the occurrence of the Lindley paradox, let us
consider the simple case where $\phi_{0}=0$ and $\sigma=1$, where 
we have (see \ref{a:pd} for the general case)
\begin{align}
\bar{z}_{0}&=p\left( \phi=\phi_{0} | x \right)  \notag 
\\ &= \left\{ 1+ 
\frac{1-z_{0}}{z_{0}} \frac{1}{\sqrt{1+\tau^{2}}}  
\exp \left[ \frac{x^{2} \tau^{2}}{2 \left( 1 + \tau^{2} 
\right) } \right] \right\} ^{-1},
\label{eq:2}
\end{align}
which goes to $1$ as $\tau$ goes to infinity, no matter 
the value of the outcome $x$ and of the prior probability $z_{0}$. 
In other words, when the prior distribution for the alternative hypothesis 
becomes non-informative (i.e. approaches a flat distribution), a Bayesian 
approach awards high odds to the null hypothesis even if the observed value 
is several standard deviation away from $\phi_{0}$. This is clearly in 
contrast with the predictions that any frequentist approach based on
sampling theory may provide, not speaking of common sense.
\par
The nature of the paradox, actually the lack of any paradox, 
has been extensively analyzed \cite{lin77,sha82}
and  we are not going to discuss the situation from a statistical point
of view, which would be beyond the scope of our analysis. 
We limit ourselves to notice that the disagreement between the 
two approaches is basically caused by the fact that the frequentist 
approach tests one hypothesis without reference to the other, whereas 
Bayesian analysis assess them as alternative one to each other, 
looking for the one in better agreement with the observations.
We also point out that the paradox is of great generality since 
the key point for its appearance is the use of a diffuse prior distribution. 
In the following sections, we will exhibit three situation of physical 
interest where the data analysis naturally involves the assignment of a
prior, such that a discussion about the occurrence of the paradox is
worthwhile.
\section{Phase estimation by homodyne detection}
\label{s:HD}
\begin{figure}[h!]
\includegraphics[width=0.9\columnwidth]{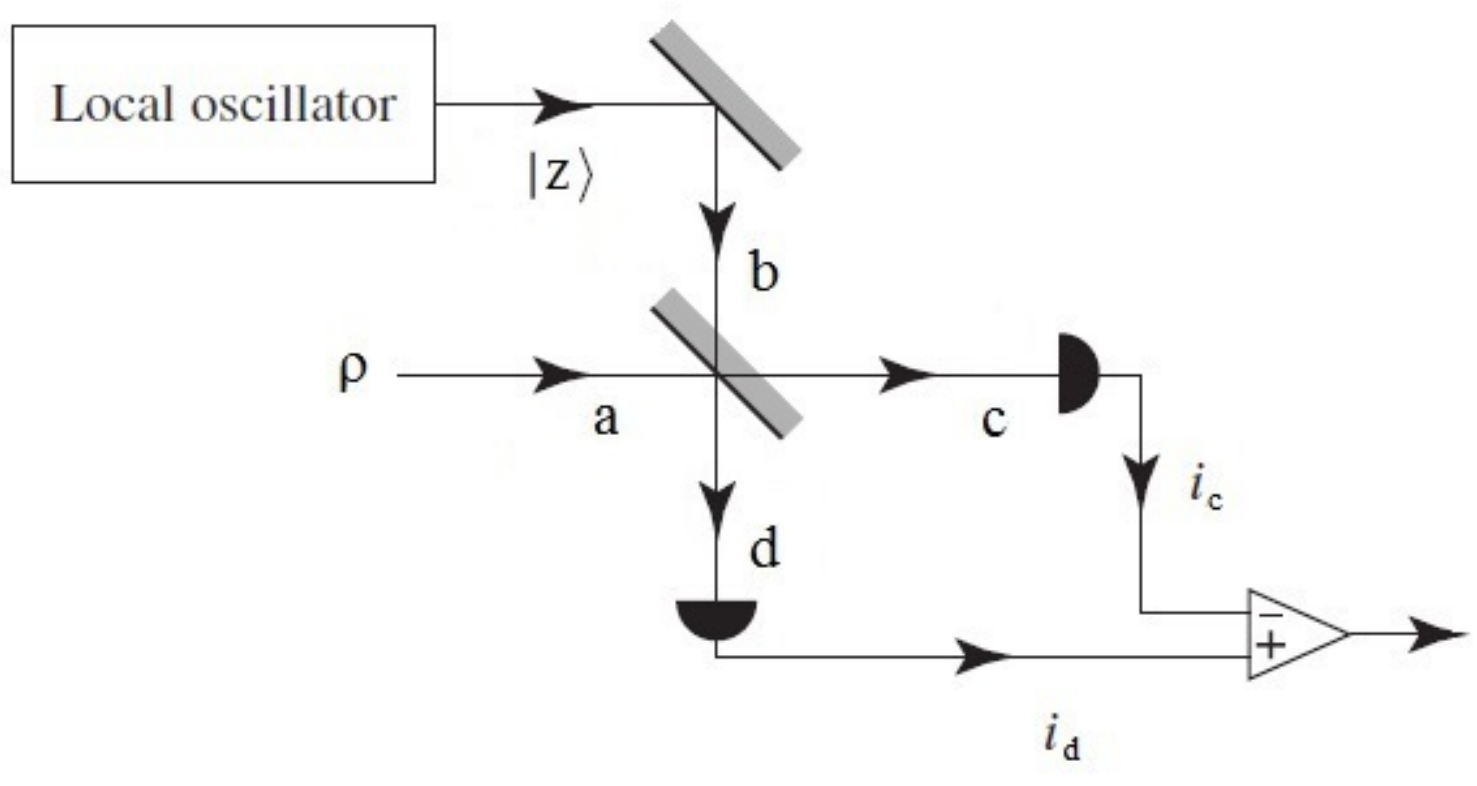}
\caption{Schematic diagram of homodyne detection scheme}
\label{fig. Homodyne Detection}
\end{figure}

Homodyne detection is a measurement scheme where the observed 
mode of the field interferes with a reference mode, thus providing 
indirect information about its phase. The device is illustrated 
schematically in
Fig. (\ref{fig. Homodyne Detection}). 
The field to be measured, prepared in an arbitrary state $\rho$ of a
single-mode $a$, is mixed with the field produced by an intense laser,
i.e. with a highly excited coherent state $\left|z\right\rangle $, with
$z=\left|z\right|e^{i\phi}\in\mathbb{C}$. Two detectors
measure the photocurrents $i_{c}$ and $i_{d}$, while a differential
amplifier determines the {\em homodyne photocurrent} $H$, i.e.  the
rescaled difference photocurrent $\hat{H}=(\hat{i}_c-\hat{i}_d)/\sqrt[2]
|z|$. In the limit of large $|z|$ it turns out that homodyne
detection provides the measurement of the quadrature operator
$\hat{\chi}_{\phi}$ given by:
\begin{equation}
\hat{\chi}_{\phi}=\frac{\hat{a}^{\dagger}e^{i\phi}+\hat{a}e^{-i\phi}}{\sqrt{2}}\,.
\label{eq:3}
\end{equation}
We point out that the device works properly only if the incoming fields 
occupy the same spatial mode, have the same frequency 
$\omega$ and are co-polarized. Besides, note that our description refers to ideal 
photodetectors with quantum efficiency equal to $1$. 
This being said, we address homodyne detection as a tool to establish 
whether the phase of a given initial field has been perturbed 
or not, e. g. due to interaction with the environment. 
In particular, we consider a squeezed state 
$\left|\alpha,\lambda\right\rangle$, with
$\alpha=\left|\alpha\right|e^{i\theta}$,
$\lambda=re^{i\varphi}\in\mathbb{C}, r>0$. 
That is, we take the impinging beam to be in a pure state given by:
\begin{equation}
\left|\psi_0\right\rangle=\left|\alpha,\lambda\right\rangle=\hat{D}(\alpha)\hat{S}(\lambda)\left|0\right\rangle \, ,
\label{eq:4}
\end{equation}
where 
$\hat{S}(\lambda)=\exp(-\frac{\lambda}{2}\hat{a}^{\dagger2}+\frac{\lambda^*}{2}\hat{a}^2)$
and $\hat{D}(\alpha)=\exp(-\alpha\hat{a}^{\dagger}+\alpha^*\hat{a})$ are
unitary operators, namely the {\em squeezing operator} and the {\em
displacement operator}, whose action in the phase space defined by the
observables $\hat{\chi}_{\phi}$ and $\hat{\chi}_{\phi+\frac{\pi}{2}}$
is, respectively, an hyperbolic rotation and a translation. Besides, we
assume that the interaction with the environment results in a phase
shift expressed by the evolution operator
$\hat{U}_I=\exp(-i\hat{a}^{\dagger}\hat{a}\beta)$, 
with $\beta\in [0,2\pi]$.  The final state $\left|\psi\right\rangle$,
impinging on the semi-reflecting mirror, is readily obtained using the
BCH formulas, which yield:
\begin{equation}
\left|\psi\right\rangle=\hat{U}_I\left|\psi_0\right\rangle=\exp(-i\hat{a}^{\dagger}\hat{a}\beta)\left|\alpha,\lambda\right\rangle
=\left|\alpha e^{-i\beta},\lambda e^{-2i\beta}\right\rangle .
\label{eq:5}
\end{equation}
The distribution $P_{\alpha,\lambda}(\chi_{\phi})$ 
of the outcomes of repeated measurements of $\hat{\chi}_{\phi}$ 
performed on radiation prepared in state $\left|\psi\right\rangle$ 
is a Gaussian with expectation value $\bar{\chi}_{\phi}$ and 
variance $\sigma^2_\chi$ given by:
\begin{equation}
\bar{\chi}_{\phi}=\left\langle\psi\left|\hat{\chi}_{\phi}\right|\psi\right\rangle
=\frac{1}{\sqrt2}\left(\alpha e^{-i(\phi+\beta)}+\alpha^* e^{i(\phi+\beta)}\right) \, ,
\label{eq:6}
\end{equation}
\begin{align}
\sigma^2_\chi &=\left\langle\psi\left|\hat{\chi}_{\phi}^2\right|\psi\right\rangle-
\left\langle\psi\left|\hat{\chi}_{\phi}\right|\psi\right\rangle^2=
\notag \\
&=\frac{1}{2}\left[e^{2r}\sin^2\left(\phi+\beta-\frac{\varphi}{2}\right)+e^{-2r}\cos^2\left(\phi+\beta-\frac{\varphi}{2}\right)\right] \, .
\label{eq:7}
\end{align}
The above equations lead us to conclude that we shall check whether the outcomes 
of measurements by means of homodyne detection are
consistent with the null hypothesis $\beta=\beta_{0}=0$ in order to 
establish whether the initial state 
$\left|\psi_0\right\rangle$ has been perturbed or not. Without loss 
of generality, we focus on a specific example and set $\theta=0$,  $\varphi=0$, 
corresponding to an input amplitude-squeezed state.
We may also fix the quadrature component to be measured: for the sake of 
simplicity we take $\phi=0$. In these conditions, 
Eqs (\ref{eq:6}, \ref{eq:7}) becomes
\begin{align}
\bar{\chi}_{0}\left(\beta\right)&=|\alpha| \sqrt{2}\cos\beta=\bar{\chi}_{0}\left(\beta\right) ,
\label{eq:8} \\
\sigma_{\chi}^{2}\left(\beta\right)&=\frac{1}{2}[e^{2r}\sin^{2}\beta+e^{-2r}\cos^{2} \beta ]\,.
\label{eq:9}
\end{align}
\par
Let us now suppose that a run returns an outcome
$q=\bar{\chi}_{0}+t\cdot\sigma_{\chi}$, with $t \in \mathbb{R}$. In
agreement with a standard sampling-theory test based on the value of $t$
we shall conclude that if $q$ is several $\sigma_{\chi}$ from
$\bar{\chi}_{0}$ the null hypothesis is to be rejected: we have to say that a phase shift
occurred and the state has been indeed perturbed.  In order to provide a
Bayesian assessment for $H_{0}$ we set a value for the prior
probability, say $z_{0}=0.9$, and distribute uniformly the remainder
probability $1-z_{0}$ over the  interval $I=\left[-\pi,\pi\right]$,
where the phase may take values. On one hand, we are allowed to set such
a high value for $z_{0}$ under the assumption that we are leading an
experiment in optimal conditions, any source of noise being taken into
account. On the other hand, the choice of a flat prior should not be
surprising since we have no idea about the dynamics of a possible
interaction: we are equally (un)expected to measure a small or a large
value for the phase shift. Eventually, the posterior probability
$\bar{z}_{0}$ that $\beta=\beta_{0}=0$ is is evaluated by the Bayes
theorem:
\begin{equation}
\bar{z}_{0} =\left\{ 1+\frac{1-z_{0}}{z_{0}}\frac{p\left(q|\beta\neq\beta_{0}\right)}{p\left(q|\beta=\beta_{0}\right)}\right\} ^{-1},
\label{eq:10} 
\end{equation}
where
\begin{equation}
p\left(q|\beta=\beta_{0}\right) =\frac{1}{\sqrt{2\pi\sigma_{\chi}^{2}\left(0\right)}}\exp\left[-\frac{q^{2}}{2\sigma_{\chi}^{2}\left(0\right)}\right] ,
\label{eq:11}
\end{equation}
\begin{equation}
p\left(q|\beta\neq\beta_{0}\right)=\int_{I}d\beta\:\frac{1}{\sqrt{2\pi\sigma_{\chi}^{2}\left(\beta\right)}}\exp\left[-\frac{\left(q-|\alpha| \sqrt{2}\cos\beta\right)^{2}}{2\sigma_{\chi}^{2}\left(\beta\right)}\right] .
\label{eq:12}
\end{equation}
Note that the value of $\bar{z}_{0}$ only depends on the value $q$ of
the outcome: we expect that the bigger is the difference
$q-\bar{\chi}_{0}$, the smaller the posterior probability $\bar{z}_{0}$.
In order to give a numerical estimate let us set $|\alpha|=10$ and
$r=1$, i.e. we consider a squeezed state with one hundred photons, where
only about $1\%$ of the total energy is employed in squeezing. Recalling
that the null hypothesis corresponds to the condition
$\beta=\beta_{0}=0$ we get $\bar{\chi}_{0}\left(0\right)=14.14$ and
$\sigma_{\chi}^{2}\left(0\right)=0.07$ .  By explicit calculations one can check that
for $\left|q-\bar{\chi}_{0}\left(0\right)\right|\leq4\sigma_{\chi}\left(0\right)$ the posterior probability is
still over $50\%$: a Bayesian approach leads us to conclude that no
phase perturbation actually occurred. That is, if 
$2\sigma_{\chi}\left(0\right)\leq\left|q-\bar{\chi}_{0}\left(0\right)\right|\leq4\sigma_{\chi}\left(0\right)$ 
the Bayesian approach provides
conflicting evidence with respect to the values obtained by a sampling
test: we are in presence of the paradox.  \par
The key point, in the present case as well as in the situation
illustrated by Lindley, is that the prior is taken to be flat. To this
regard we remark that despite this assumption is somehow consistent from
a physical perspective, in practice it turns out that outcomes being
several $\sigma_{\chi}\left(0\right)$ from the expectation value are associated to
statistical fluctuations and the null hypothesis keeps on overwhelming
any alternative. For this reason we shall rather conclude that the
Bayesian approach is somehow misleading if a scarce information about
the prior distributions is available, whereas a sampling-theory test
better accounts what is really happening.
\par
Since our main goal is to address the occurrence of the Lindley paradox
in  interferometry we will not go further in the discussion of this
case. We only notice that here we have a physical example where
the paradox unavoidably arises, the assumptions being consistent with
the dynamics of the system. From a practical point of view, we notice
that the characterization and the calibration of homodyne detectors is one of
the building blocks for the robust implementation of quantum tomography
of the radiation field \cite{reh04}.
\section{Phase Estimation with Mach-Zehnder Interferometry}
\label{s:MZI}
\begin{figure}[h!]
\includegraphics[width=0.9\columnwidth]{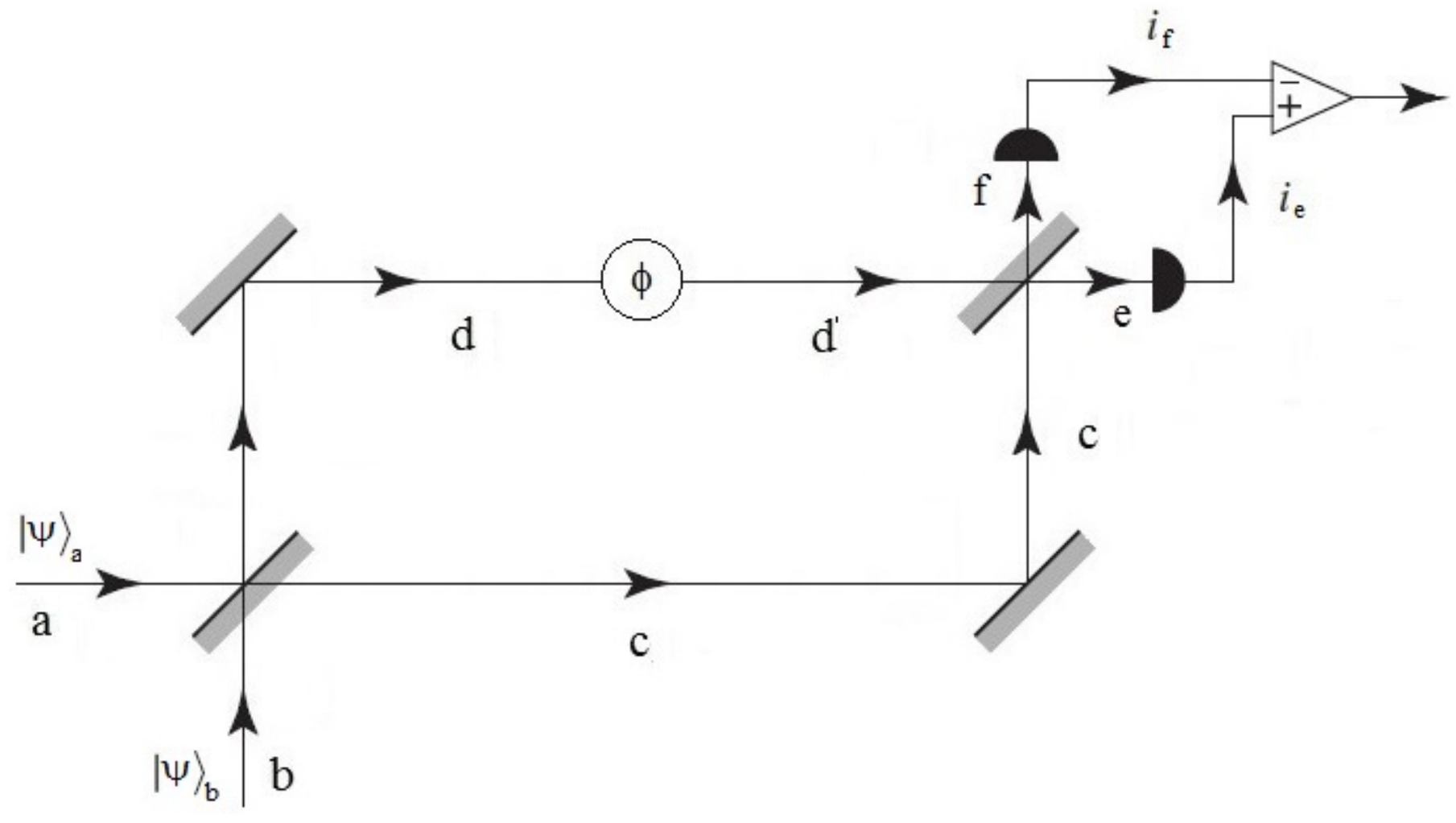}
\caption{Schematic diagram of a Mach-Zehnder interferometer.}
\label{fig:MZI}
\end{figure}

In the present section we consider the linear MZ interferometer, that
allows one to monitor a classical phase parameter via photon counting
measurements. A schematic diagram of the detector is shown in Fig.
\ref{fig:MZI}.  
\par
Two modes of the radiation field enters the interferometer through a
first beam splitter, then a phase difference $\phi$ is inserted between
the beams and, finally, the two fields exit through a second beam
splitter. The number of photons in each mode is detected by means of two
photodetectors of quantum efficiency $\eta$ that measure the
photocurrents $i_{f}$ and $i_{e}$; a differential amplifier eventually
determines the difference photocurrent, whose possible values correspond
to the eigenvalues of the Hermitian operator
\begin{equation}
\hat{D}(\phi)=\eta\left(\hat{f}^{\dagger}\hat{f}-\hat{e}^{\dagger}\hat{e}\right)\,
\label{eq:13}
\end{equation}
which is a normalized version of the integrated photocurrent. 
It can be expressed in terms of the input fields as follows:
\begin{equation}
\hat{D}(\phi)=\eta\left[\left(\hat{a}^{\dagger}\hat{a}-
\hat{b}^{\dagger}\hat{b}\right)\cos\phi+i
\left(\hat{b}^{\dagger}\hat{a}-\hat{a}^{\dagger}
\hat{b}\right)\sin\phi\right] .
\label{eq:14}
\end{equation}
From the above equation it is clear that measurements of the difference
photocurrent at the output allows one to assess variations of the phase
difference $\phi$. In particular, from a linear error propagation theory
one obtains the following expression for the sensitivity $\Delta\phi$ of
the device:
\begin{align}
\Delta\phi=\sqrt{\left\langle \hat{D}^{2}(\phi)\right\rangle
-\left\langle \hat{D}(\phi)\right\rangle
^{2}}\;\left|\frac{\partial}{\partial\phi}\left\langle
\hat{D}(\phi)\right\rangle \right|_{\phi=\phi_{0}}^{-1}, \label{eq:15}
\end{align}
which depends from the initial value $\phi_{0}$ and the expectation
value $\left\langle \hat{D}(\phi)\right\rangle$. If any perturbations
changes the length of the two arms of the interferometer, the optical 
paths covered by the light beams is altered. 
As a consequence, the phase difference
$\phi$ will vary and so will do the outcomes of measurements of the
operator $\hat{D}$: once the system has been set in a proper initial
working point $\phi_{0}$ an outcome $\phi\neq\phi_{0}$ shall be
associated to a displacement. Hence, the data analysis
consists of assessing the probability that an outcome $d$ confirms the
null hypothesis $H_{0}$, corresponding to no phase variation, or the
alternative hypothesis $H_{1}$ that $\phi\neq\phi_{0}$. In this context,
we discuss the possibility of coming across the Lindley paradox,
paying particular attention to the conditions in which it can arise.
\subsection{Coherent light interferometry}
At first, we consider a classical-like MZ interferometer fed by an input
state of the form $\left|\psi\right\rangle =\left|\alpha\right\rangle
\left|0\right\rangle$ where
$\alpha=\left|\alpha\right|e^{i\theta}\in\mathbb{C}$. In this case, the
outcomes $d$ of measurements of the operator $\hat{D}$ are distributed
according to a Skellam distribution \cite{ske46}:
\begin{equation}
p_{\phi}(d)=e^{-\eta\left|\alpha\right|^{2}}
\left(\frac{1+\cos\phi}{1-\cos\phi}\right)^{\frac{d}{2}}
I_{\left|d\right|}\left(\eta\left|\alpha\right|^{2}\sin\phi\right),
\label{eq:16} \end{equation}
where $I_k(z)$ is the modified Bessel function of the first kind.
The mean value $\bar{d}_{\phi}$ and variance 
$\sigma_{d}^{2}$ of the above distribution 
are given by 
\begin{equation}
\bar{d}_{\phi}=\left\langle \hat{D}
(\phi)\right\rangle=\eta\left|\alpha\right|^{2}\cos\phi\,,
\label{eq:17}
\end{equation}
\begin{equation}
\sigma_{d}^{2}= \Delta^{2}\hat{D}=\eta\left|\alpha\right|^{2}.
\label{eq:18}
\end{equation}
This result may be easily understood recalling that the number of
photons contained in a coherent state follows a Poissonian statistics and
that the Skellam distribution provides the probability density of the
difference between two Poisson variates.  It is easy to see that for the
given input state $\left|\psi\right\rangle$ the minimum value of
$\Delta\phi$ is obtained for $\phi_{0}=\frac{\pi}{2}$. Therefore, we
assume the interferometer to be set in this optimal working point
(corresponding to a null mean value $\bar{d}_{\frac{\pi}{2}}=0$).
\par
We now suppose that a trial results in the outcome $d$ of the difference
photocurrent, that we represent as
$d=\bar{d}_{\frac{\pi}{2}}+t\cdot\sigma_{d}=t\cdot\sigma_{d}$, with
$t\in\mathbb{R}$. According to a standard sampling-theory test $d$ is to
be associated to a displacement induced by a perturbation 
if its value is several
$\sigma_{d}$ from $0$. In this event we shall reject the null hypothesis
$H_{0}$. On the other side, in order to give a Bayesian assessment to
$H_{0}$ we assign a probability $z_{0}$ to the null hypothesis and
distribute the remainder over the interval $I=\left[0,\pi\right]$ where
we expect the phase to vary. Assuming to deal with rare events we
take $z_{0}=0.99$, i. e. we assume that almost only $1$
event out of $100$ is to be associated to an incoming perturbation 
rather than to
a random error. With respect to the prior distribution for the
alternative hypothesis we initially consider a situation where no
information about the signal is available so that a
flat prior results a proper choice. At this point we can compute the
posterior probability $\bar{z}_{0}$ given $\phi=\phi_{0}=\frac{\pi}{2}$
by Eq. (\ref{eq:10}), where $\beta$ is to be substituted with $\phi$ and
the probability densities take the following expressions:
\begin{align}
p\left(d|\phi=\phi_{0}\right) & =\int_{I}d\phi\:
p_{\phi}\left(d\right)\delta\left(\phi-\phi_{0}\right) \nonumber \\ 
& =e^{-\eta\left|\alpha\right|^{2}}
I_{\left|d\right|}(\eta\left|\alpha\right|^{2}),
\label{eq:19}
\end{align}
\begin{equation}
p\left(d|\phi\neq\phi_{0}\right)=\int_{I}d\phi\: p_{\phi}\left(d\right).
\label{eq:20}
\end{equation}
\begin{figure}[h!]
\includegraphics[width=0.48\columnwidth ]{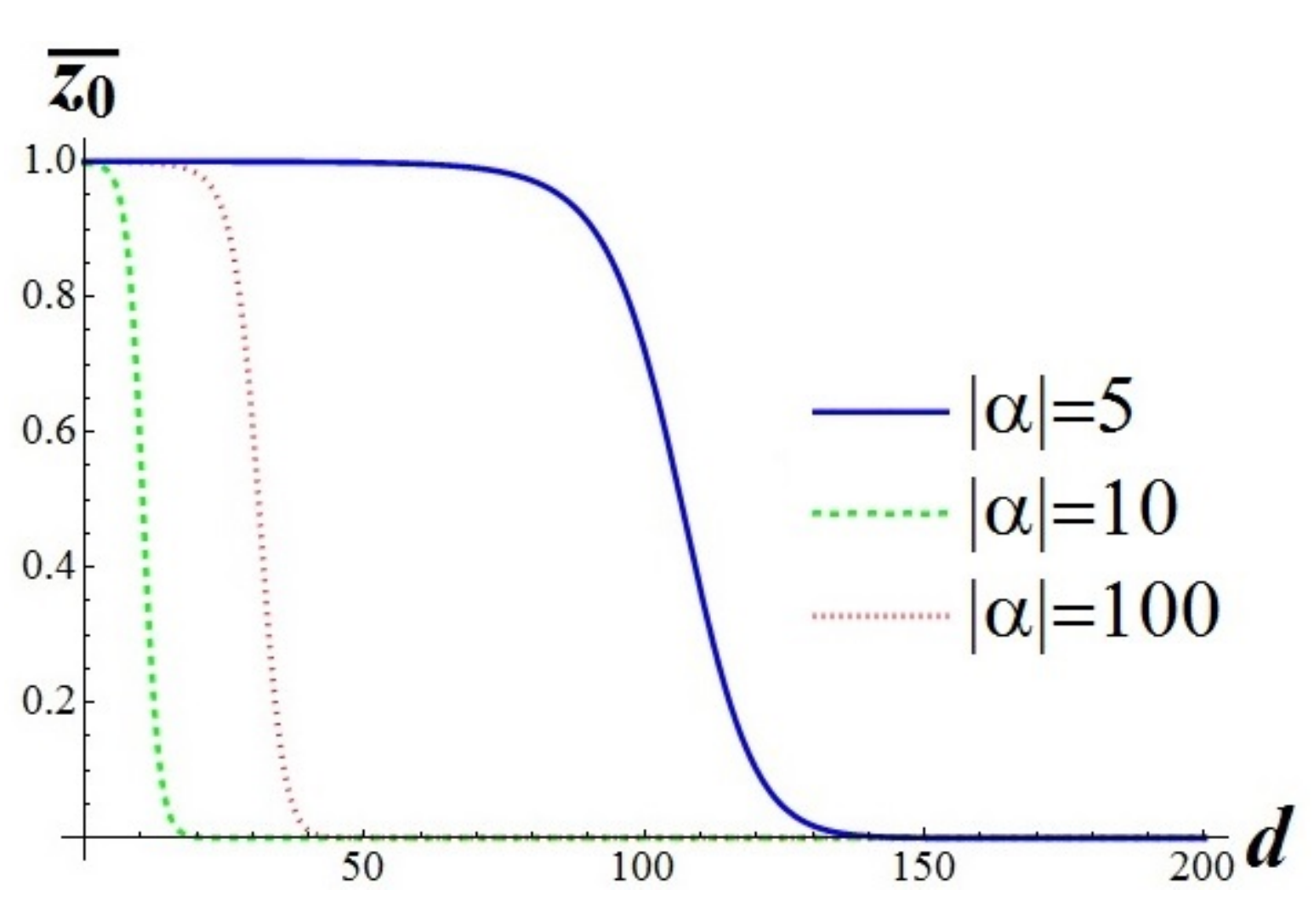}
\includegraphics[width=0.48\columnwidth ]{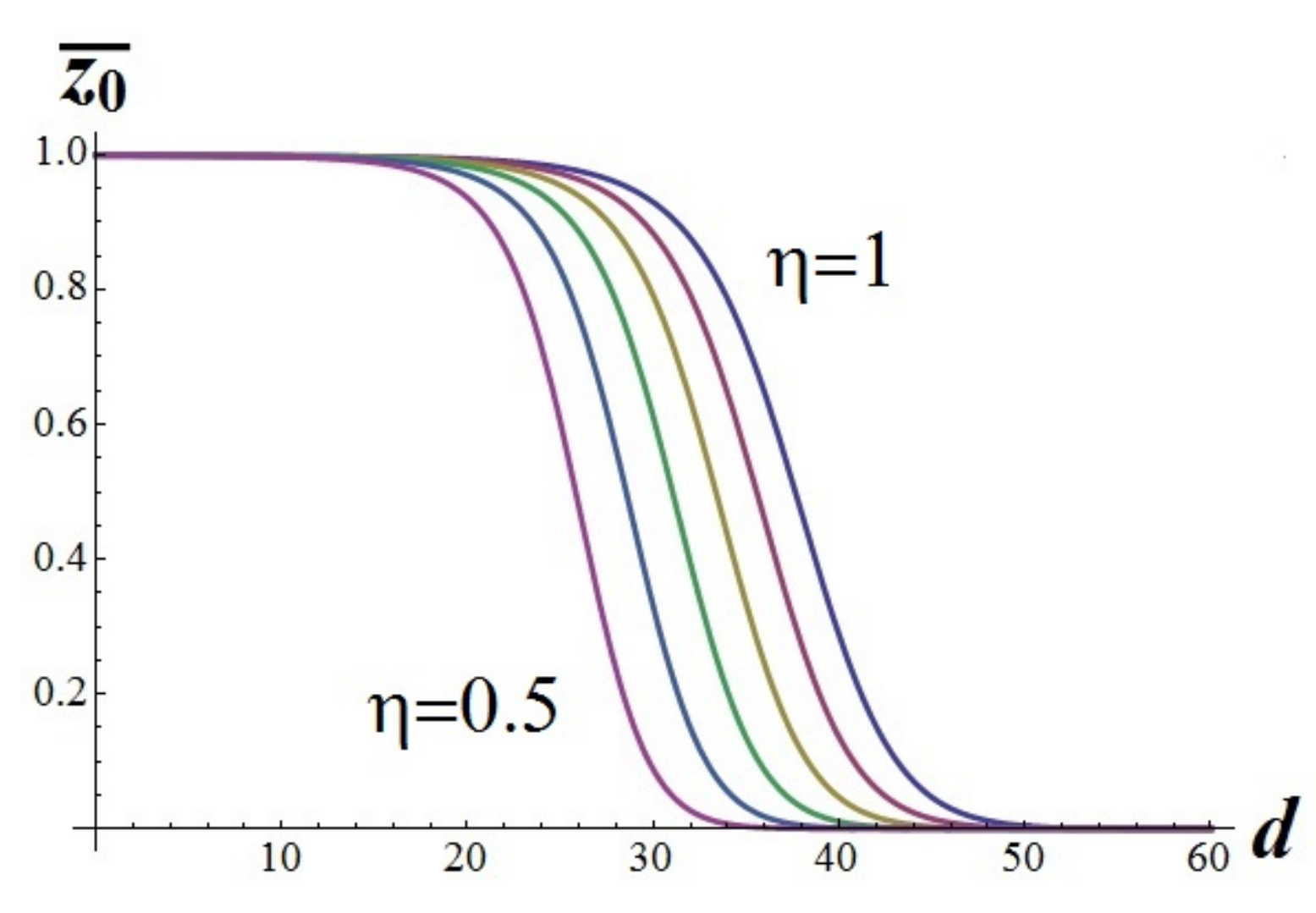}
\caption{Bayesian posterior probability $\bar{z}_{0}\left(d\right)$ in
coherent light interferometry with diffuse prior. 
Left: we show $\bar{z}_0 (d)$ as a
function of $d$ for different values of $\left|\alpha\right|$ and fixed
quantum efficiency $\eta=0.7$. Right: $\bar{z}_0 (d)$ as a function of
$d$ for different values of $\eta$ and fixed photon number
$\bar{n}=100$. Given the distance
$t=({d-\bar{d}_{\frac{\pi}{2}}})/{\sigma_{d}}$ between the outcome and
the expectation value the probability $\bar{z}_{0}\left(d\right)$ varies
slowly with $\eta$.} 
\label{f3coh}
\end{figure}
\par
The behavior of $\bar{z}_{0}\left(d\right)$ with respect to the number
of photons injected and the quantum efficiency of the detectors is
reported in Fig. \ref{f3coh}. 
We note that the trend is the same for different $\bar{n}$ and
$\eta$: the presence of non unit quantum efficiency in the detection process does not
change the shape of $\bar{z}_{0}\left(d\right)$ and the same goes for
the light intensity.  In facts, the overall effect is to reduce the
posterior probabilities, i. e. given an outcome $d$ and
$\eta_{1}<\eta_{2}$ we have
$\bar{z}_{0,1}\left(d\right)<\bar{z}_{0,2}\left(d\right)$, i.e. 
we get a lower value of
$\bar{z}_{0}\left(d\right)$ when we lower down the efficiency of the
apparatus. 
\par
That being said, for a given number of photons impinging on the device
and efficiency of the photocounters, we can identify three different
regions corresponding to different values of the outcome. In the range
$0\leq d\leq2\sigma_{d}$ the posterior probability is approximately $1$,
in agreement with a sampling-theory approach.  In this case the phase
shift is so little that its origin is much more likely to be a random
fluctuation rather than an actual perturbation. 
If $d \geq 4\sigma_{d}$ the
posterior probability is approximately $0$: both the Bayesian and the
Frequentist approach affirm that the phase shift has been induced by a
perturbation. At last, $\bar{z}_{0}\left(d\right)$ falls from $1$ to $0$ as the
value of the outcome increases within the interval $2\sigma_{d}\leq
d\leq4\sigma_{d}$. Since a frequentist analysis shall provide a value of
$\bar{z}_{0}$ below $0.05$ we have that in this region we are in
presence of the paradox.
\par
On the other hand, one may find uneasy to say that an outcome four times
$\sigma_{d}$ far from the mean value is to be associated to statistical
fluctuations. In facts, we can describe what is happening for
$2\sigma_{d}\leq d\leq4\sigma_{d}$ as follows. Our
measurement $d$ of the difference photocurrent gave a random value that
was manifestly inconsistent with the expected value in absence of
interaction; however, as a consequence of the prior assumed we are led
to conclude that it would be even more improbable to observe that
specific outcome were it the result of an interaction. That is, 
what makes the paradox arise is that
the prior asks us to regard at every alternative possibility as an
improbable coincidence. For this reason, in a real experiment an outcome
in the range $2\sigma_{d}\leq d\leq4\sigma_{d}$ should be interpreted as 
evidence of detection.
\par
\begin{figure}[h!]
\includegraphics[width=0.48\columnwidth]{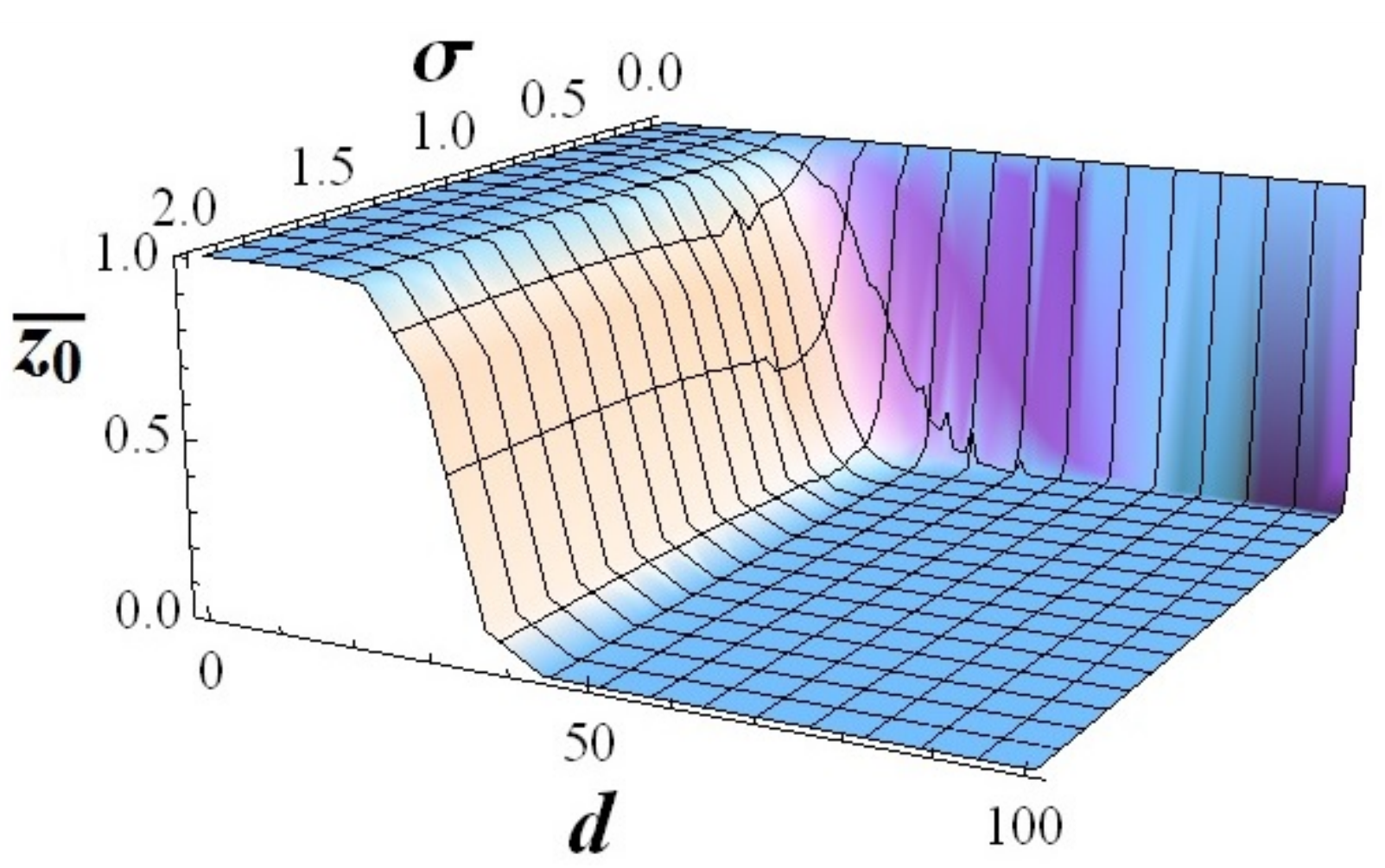}
\includegraphics[width=0.48\columnwidth]{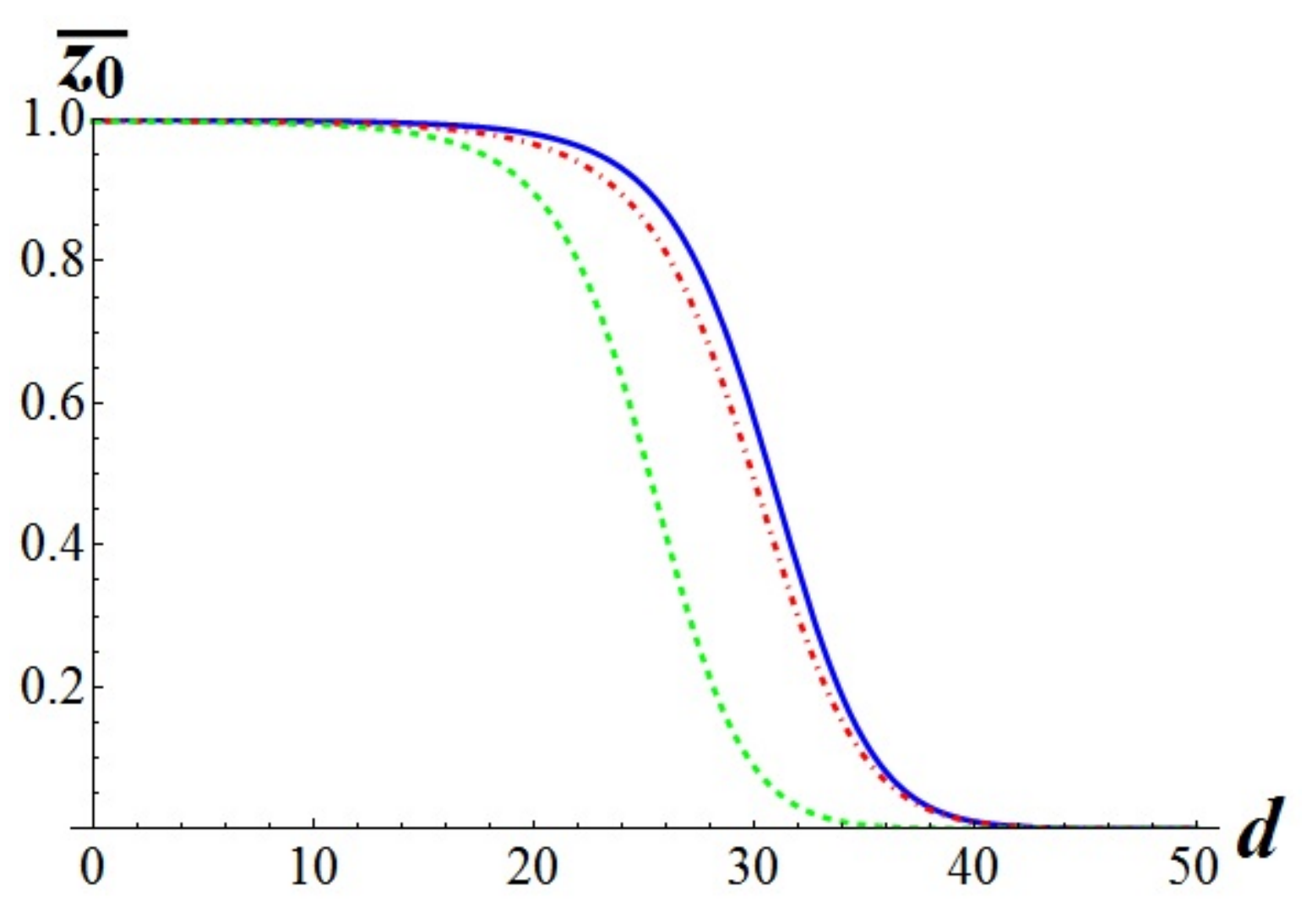}
\caption{Posterior probability $\bar{z}_{0}$ in coherent light
interferometry with wrapped normal prior. Left: $\bar z_0 (d)$
as function of $d$ and
$\sigma$ with $\eta=0.7$. The coherent amplitude is $\alpha=10$ so that
$\bar{n}=100$ and $\sigma_{d}^{2}=70$. An actual disagreement with respect to a
frequentist approach occurs for $0.2\leq\sigma \cap 2\sigma_{d}\leq
d\leq4\sigma_{d}$.
Right: $\bar z_0 (d)$ as a function of $d$. 
The continuous blue line and the
dot-dashed red line are obtained for $\eta=0.7$ and $\sigma=1;
\sigma=0,5$ respectively.  The dashed green line refers to the case
$\eta=0.5$ and $\sigma=0,5$.
} \label{fig4}
\end{figure}

In approaching the most realistic description, we now discuss the
occurrence of the paradox when a less diffuse prior is assumed. 
Here we consider a Gaussian-like distribution
as a prior in order to account for our belief that a 
perturbation results in a small phase-shift around
the expected value. In particular, we consider the so-called {\em
wrapped normal} distribution, defined as follows:
\begin{equation}
g_{\sigma,\phi_{0}}\left(\phi\right)=\frac{1}{\sqrt{2\pi\sigma^{2}}}\sum_{k=-\infty}^{\infty}\exp
\left[-\frac{\left(\phi-\phi_{0}+k\pi\right)^{2}}{2\sigma^{2}}\right],
\label{eq:21}
\end{equation}
The wrapped distribution is normalized on the interval $I$ and is
characterized by two parameters, $\phi_{0}$ and $\sigma^{2}$, which
correspond to the expectation value and variance of a normal
distribution. In recalling that the mean value of the outcomes is
$\bar{d}_{\frac{\pi}{2}}=0$ we impose $\phi_{0}=\frac{\pi}{2}$ but we
let the variance of the prior free to vary in order to assess the
relevance of our knowledge about the system with respect to the
statistics.  Note that since $\bar{z}_{0}$ is inversely proportional to
the ratio
$p\left(d|\phi\neq\phi_{0}\right)/p\left(d|\phi=\phi_{0}\right)$, which
increases as the variance of the prior assumed for the alternative
hypothesis decreases, we expect that the sharper is the the prior
considered the higher will be the value of $\bar{z}_{0}$. For this
reason we may expect the Lindley paradox not to occur when the variance
$\sigma$ of $g_{\sigma,\phi_{0}}\left(\phi\right)$ is taken to be little
with respect to the one of the Skellam distribution, given by Eq.
(\ref{eq:17}).  We report in Fig. \ref{fig4} the results obtained substituting
equation (\ref{eq:16}) with the proper expression 
for the prior (\ref{eq:21}), so that:
\begin{equation}
p\left(d|\phi\neq\phi_{0}\right)=\int_{I}d\phi\:
p_{\phi}\left(d\right)g_{\sigma,\frac{\pi}{2}}\left(\phi\right).
\label{eq:22}
\end{equation}
\par
At first, we see that if the variance of the prior is taken too small
then the posterior probability does not evolve (with reference to the 
plotted range for d): we get
$\bar{z}_{0}=z_{0}$ independently from the outcome $d$. From a
mathematical perspective, this is due to the fact that the ratio
$p\left(d|\phi=\phi_{0}\right)/p\left(d|\phi\neq\phi_{0}\right)$
appearing in the expression of $\bar{z}_{0}$ tends to unity as
$p\left(d|\phi\neq\phi_{0}\right)$ tends to a delta
$\delta\left(\phi-\phi_{0}\right)$.  In facts, we are assuming a sharp
prior centered in the mean value both for the null hypothesis and the
alternative hypothesis. That is, we are somehow stating that $H_{0}$ and
$H_{1}$ are equivalent as both of them, according to our prior models,
lead to a measurement of $d$ to be associated to the outcome
$d=\bar{d}_{\frac{\pi}{2}}$. Therefore, after a trial we cannot conclude
anything else but that it is to be associated to a measurement of the
mean value. Similarly, an evaluation of the posterior probability for
the alternative hypothesis $p\left(d|\phi\neq\phi_{0}\right)$ yields
$\overline{{1-z}_{0}}=1-z_{0}$. This last remark makes it clear that 
we are providing too much prior information and Bayesian
inference can only confirm what we already know, without the possibility
of improving our knowledge about the system. Conversely, if the prior
distribution is diffuse, i. e.  $\sigma>1$, we recover the case of
flat prior examined above, where our knowledge is poor and, in turn, for
$2\sigma_{d}\leq d\leq4\sigma_{d}$ the paradox is again observed.
\par
The most interesting region is $0.2\leq\sigma\leq1$. A comparison
between the values of $\bar{z}_{0}$ obtained employing a widespread
prior and a sharper one, see the right panel of Fig. \ref{fig4}, 
shows that the latter distribution awards less
probability to the null hypothesis, as we were expected to. However, the
weight of the prior is not sufficient to avoid the paradox, which still
occurs if $2\sigma_{d}\leq d\leq4\sigma_{d}$ even for small values of
the variance of the wrapped normal distribution.  Fig. \ref{fig4} 
also suggests that if the interferometer
has poor detectors then the odds in favor of the alternative hypothesis
will be lower.  Nevertheless, in recalling that $\sigma_{d}$ decreases with
$\eta$ it is possible to see that the net effect remains unchanged.
\subsection{Squeezed light interferometry}
Let us now consider an input state of the form $\left|\psi\right\rangle
=\left|\alpha\right\rangle \left|\lambda\right\rangle$, where
$\alpha=\left|\alpha\right|e^{i\theta},\lambda=re^{i\varphi}\in\mathbb{C}$.
That is, we replace the vacuum state $\left|0\right\rangle$ in mode $b$
with a squeezed-vacuum state $\left|\lambda\right\rangle$, containing
$\bar{n}=\sinh^2 (r)$ photons. The advantages gained by squeezing
enhancement have been widely studied in literature. Here, our aim is 
to assess the
convenience of employing squeezed light from a statistical perspective.
That is, we address the possibility of getting by the Lindley paradox
when the injected beam is previously squeezed.
\par
Similarly to the previous case we will evaluate the posterior
probability using a flat prior, at first, and then a peaked
distribution.  At the same time, we will pay attention to the
contribution of the quantum efficiency.  The difference with respect to
the classical scheme seats in the distribution of the outcomes
$p_{\alpha,\lambda,\phi}(d)$, which turns out to be difficult to be
computed. However, since the expectation value
$\bar{d}_{\phi}$ and the variance $\sigma_{d}^{2}$ may be 
calculated exactly we employ a Gaussian distribution 
for the outcomes as a first-order approximation:
\begin{equation}
p_{\alpha,\lambda,\phi}(d)=\frac{1}{\sqrt{2\pi\sigma_d^2}} 
\, e^{-[d-\eta \cos\phi(|\alpha|^2-
\sinh^2 r)]^2/2\sigma_d^2}\,,
\label{eq:23}
\end{equation}
where the variance is a function of the initial working point $\phi_{0}$ and of 
the measured phase shift $\phi$. 
Similarly to what happened in the previous case, $\sigma_{d}^{2}$ takes the 
minimum value at $\phi_{0}=\pi/2$, namely:
\begin{align}
\sigma_{d}^{2}=&\frac{1}{2}\eta\,(1-\eta)(|\alpha|^{2}+
\sinh^{2} r ) \nonumber\\
&+\eta^2 \Bigg\{|\alpha|^{2}+2\sinh^{2} r \cosh^{2} r \cos^2 \phi 
\label{eq:24} \\
&+\sin^2 \phi\left[\sinh^{2} r (1+2\alpha^2)-
2\alpha^2\sinh r \cosh r \right]\Bigg\}
\nonumber
\end{align}
whereas the expectation value results $\bar{d}_{\frac{\pi}{2}}=0$.
\par
\begin{figure}[t!]
\includegraphics[width=0.48\columnwidth]{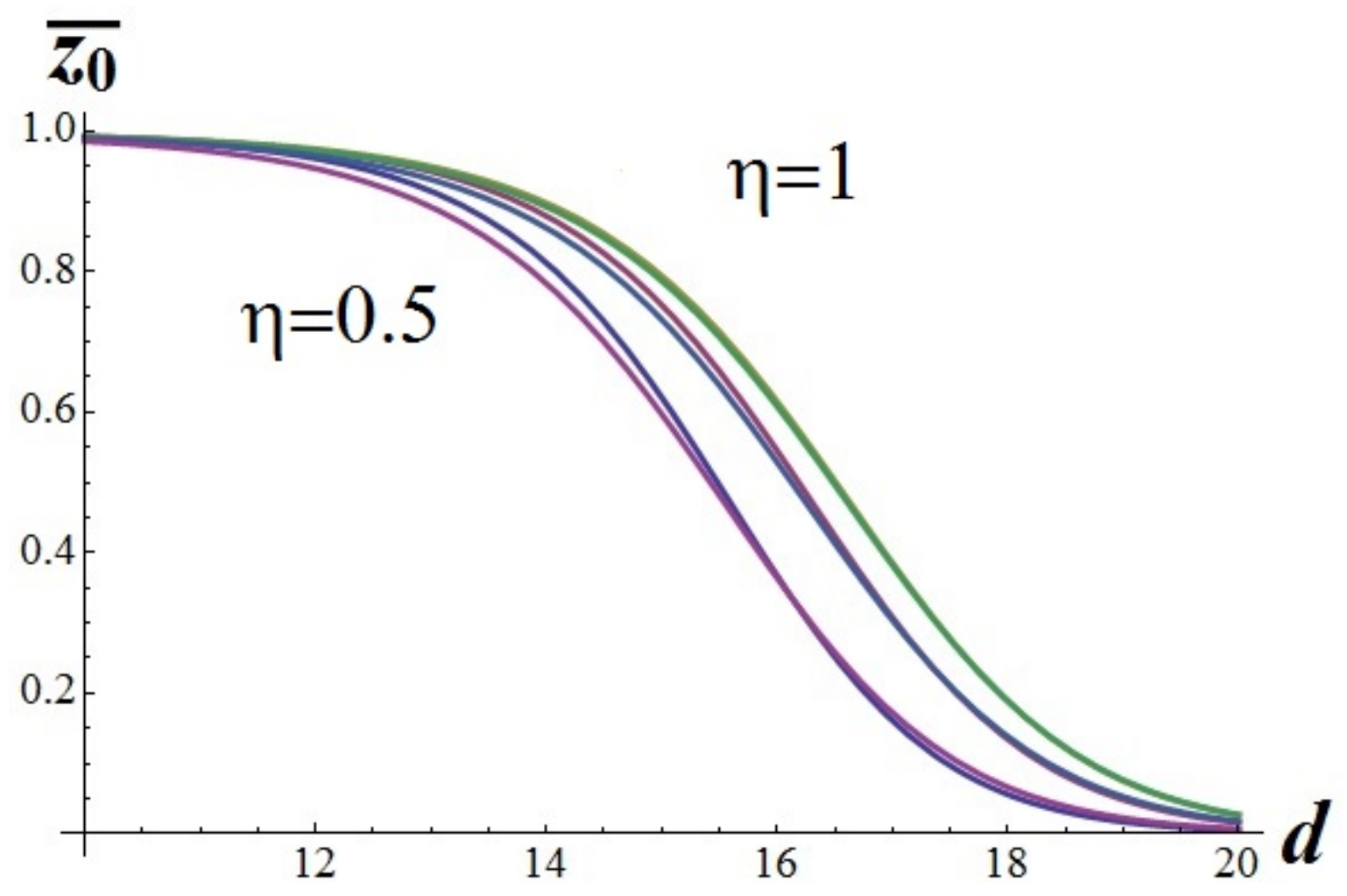}
\includegraphics[width=0.48\columnwidth]{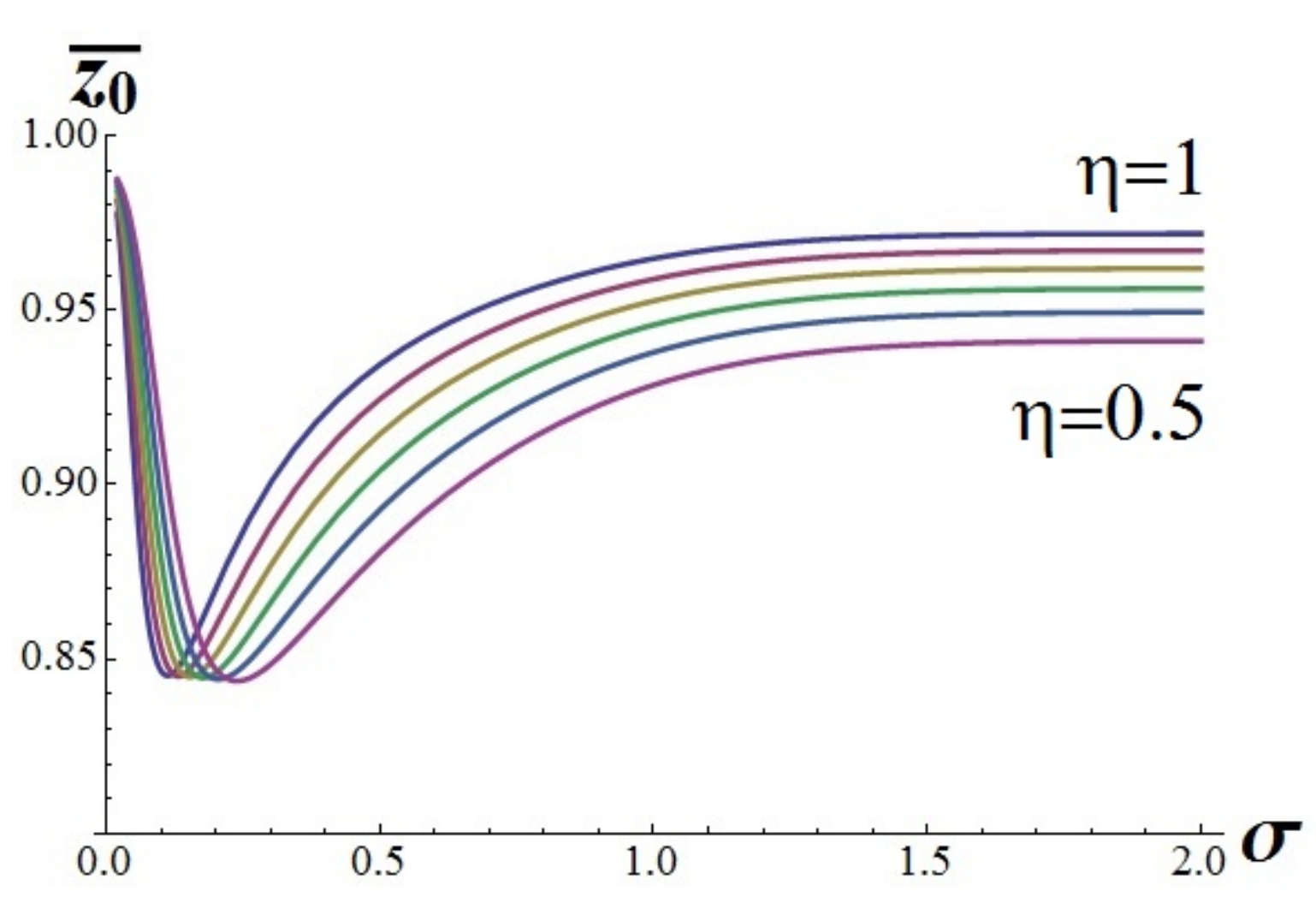}
\caption{Bayesian posterior probability $\bar z_0 (d)$ in squeezed light
interferometry. Left: $\bar z_0 (d)$ as a function of $d$ for efficiency 
$\eta$ from $0.5$ to $1.0$ and for $\alpha=10$, $r=1$, $\varphi=\pi$. 
Right: the posterior probability for given distance
$|d-\bar{d}_\phi|=3\sigma_d$ and the same values of the state parameters (while $\phi_0=\pi/2, \phi=0$ in $\sigma_{d}$).}
\label{fig5}
\end{figure}

 
In order to test the consistency of an outcome
$d=\bar{d}_{\frac{\pi}{2}}+t\cdot\sigma_{d}=t\cdot\sigma_{d}$ with the
null hypothesis $H_0$ we pose $z_{0}=0.99$ and assume that the remainder
probability $1-z_{0}$ is uniformly distributed over the interval
$I=\left[0,\pi\right]$. Before computing the posterior probability we
point out that the best sensitivity $\Delta\phi$ is achieved for
$\varphi=\pi$ and $\left|\alpha\right| ^{2}\gg\sinh^{2}\left(r\right)$. 
The posterior probability $\bar{z}_{0}$ that $\phi=\phi_{0}$ is, then, 
obtained by means of equation (\ref{eq:10}), with $\beta$ equal to 
$\phi$, leading to
\begin{align}
p\left(d|\phi=\phi_{0}\right)&=
\int_{I}d\phi\: p_{\alpha,\lambda,\phi}
\left(d\right)\delta\left(\phi-\phi_{0}\right) 
\label{eq:25} \\
p\left(d|\phi\neq\phi_{0}\right)&=
\int_{I}d\phi\: p_{\alpha,\lambda,\phi}\left(d\right).
\label{eq:26}
\end{align}
The behavior obtained for $\bar{z}_{0}$ is very close to that 
observed for its classical counterpart, i.e. the dashed green line in
Fig. \ref{f3coh}.
Indeed, we can identify a region where $\bar{z}_{0}$ tends to unity (for
small values of $d$), a region where it is approximately zero (to the
right) and a "transition area'', where the posterior probability varies
between these two limits.  What matters most is that we are again in
presence of the Lindley paradox: if we test the hypothesis
concerning the detection of a perturbation 
by means of the Bayes' theorem we will
end up in a misleading analysis. It is possible to see from the plot of
Fig.  \ref{fig5}
that the situation does not change in presence of non ideal
photodetectors. In the same vein, when we consider input states
containing a higher number of photons and such that the optimal
conditions mentioned above are still satisfied (i. e. $\varphi=\pi$ and
$\left|\alpha\right| ^{2}\gg\sinh^{2}r$), the posterior
probability approaches zero for lower values of $d/\sigma_{d}$. However,
the contribution gained from the use of higher intensity light is not
sufficient to make it disappear.  \par At last, we take into account the
most actual case with squeezed input light and a Gaussian-like prior
distribution wherein the statistical description. The change in the
shape of $p\left(d|\phi\neq\phi_{0}\right)$ has the usual effect of
decreasing the value of $\bar{z}_{0}$, as it is illustrated in Fig.
\ref{fig5}.
Nevertheless, we have to say that neither in this conditions we can have
a consistent Bayesian statistical inference. That is, the value of the
posterior probability does not decrease, not even taking an appreciably
peaked prior, until the outcome is more than several $\sigma_{d}$ far
from the expectation value.
\section{Conclusions}
\label{s:outro}
We have addressed the occurrence of the so-called Lindley paradox in
the analysis of data coming from homodyne detection and optical
interferometry.  We found that the Lindley paradox may indeed occur.  In
particular, we have shown that the Bayesian approach is somehow
misleading if a scarce information about the prior distributions is
available, as it happens in the evaluation of the effect of an external
perturbation, and sampling-theory should be preferred.  Concerning MZ
interferometers, we have shown that Lindley paradox appears both for
coherent and squeezed signals and is present for any value of the
quantum efficiency of the involved detectors.  On the other hand, the
disagreement between Bayesian and frequentist approach is less
pronounced for increasing noise and may be softened by using a suitable,
more localized, priors.
\par
Our results, besides being of fundamental interest for interferometry, 
are of practical significance for quantum state
reconstruction, where calibration of homodyne detectors represents a
crucial step for the implementation of quantum tomography \cite{reh04}.
\section*{Acknowledgment}
This work has been supported by EU through the Collaborative 
Project QuProCS (Grant Agreement 641277) and by UniMI through 
the H2020 Transition Grant 15-6-3008000-625.
\appendix
\section{Posterior probabilities}\label{a:pd}
Consider a physical system where the quantity $X$ may be measured
and assume that the outcomes from the measurement of $X$ 
are distributed according to a normal distribution 
with an unknown mean value $\phi$ and a known variance $\sigma^2$.
We perform a measurement of $X$ and on the basis
of the outcome $x$ we want to test which of the two following
hypothesis is true
\begin{align}
&H_0: \hbox{the mean value is equal to a given value}\,\, 
\phi=\phi_0 \label{a1} \\
&H_1: \hbox{the mean value is not the value above}\,\, 
\phi\neq \phi_0 \label{a2}\,.
\end{align}
We assume to have some prior knowledge on the system, which may be
expressed in form of some {\em a priori} probabilities $p(H_0)$ and 
$p(H_1) = 1 - p(H_0)$ for the two hypothesis. The measurement of 
$X$ is, in turn, intended to upgrade our knowledge of the system.
Given the result $x$, Bayes theorem says that 
$$ p(H_k|x) p(x) = p(x|H_k) p(H_k)\,,$$ for $k=0,1$, where
$p(H_k)$ is the a priori probability introduced above, $p(x|H_k)$
is the conditional distribution of the outcomes given a hypothesis,
$p(x) = p(x|H_0)p(H_0) + p(x|H_1) p(H_1)$ is the overall
distribution of the outcomes, independently which hypothesis
is actually true, and
$ p(H_k|x)$ is the {\em a posteriori} probability of the hypothesis
$H_k$, i.e. the quantity of interest here.
Using the Bayes theorem we may evaluate the a posteriori probabilities, 
e.g. $p(H_0|x)$ is given by
\begin{align}
p(H_0|x) & = \frac{p(x|H_0) p(H_0)}{p(x)} \nonumber \\
& = \frac{p(x|H_0) p(H_0)}{p(x|H_0)p(H_0) + p(x|H_1) p(H_1)} 
\nonumber \\
& = \left[1+ \frac{p(x|H_1) p(H_1)}{p(x|H_0) p(H_0)} \right]^{-1}
\nonumber \\
& = \left[1+ \frac{1-p(H_0)}{p(H_0)} \frac{p(x|H_1)}{p(x|H_0)} \right]^{-1}
\,,
\label{a3}
\end{align}
which, upon the substitutions $p(H_0) \rightarrow z_0$ and 
$p(H_0|x) \rightarrow \bar z_0$, coincides with Eq. (\ref{eq:1}).
\par
Now, in order to evaluate explicitly the posterior distribution we
assign some prior distribution $\pi_k(\phi) \equiv p(\phi|H_k)$ 
to the mean value under the two hypothesis. According to (\ref{a1})
and (\ref{a2}) we write $$\pi_0(\phi) = \delta(\phi-\phi_0)\,$$
and assign the rest of the prior probability as a normal distribution
with variance $\tau$, i.e. $$\pi_1(\phi) = (2\pi\tau^2)^{-1/2} \exp\{-
(\phi-\phi_0)^2/2\tau^2\}\,.$$
\par
It is now straightforward to evaluate the conditional probabilities
as $$ p(x|H_k) = \int\! d\phi\, \pi_k(\phi)\, p(x|\phi)\,,$$
which leads to
\begin{align}
p(x|H_0) &= (2\pi \sigma^2)^{-1/2}\, e^{-(\phi-\phi_0)^2/2\sigma^2}
\\
p(x|H_1) &= [2\pi (\sigma^2+\tau^2)]^{-1/2} 
\,e^{-(\phi-\phi_0)^2/2(\sigma^2+\tau^2)}\,.
\end{align}
Overall, Eq. (\ref{a3}) rewrites as
\begin{align}
p(H_0|x) = \Bigg[1 + & 
\frac{1-p(H_0)}{p(H_0)} \sqrt{\frac{\sigma^2}{\sigma^2+\tau^2}} 
\nonumber \\ 
\times &
\exp\left(\frac{\tau^2}{2\sigma^2}
\frac{(x-\phi_0)^2}{\sigma^2+\tau^2}\right)
\Bigg]^{-1}\,.
\end{align}
Upon substituting $\phi_0=0$ and $\sigma=1$ we arrive at Eq. (\ref{eq:2}).

\end{document}